\documentclass[twocolumn,tighten,trackchanges,preprint2]{aastex631}
\hypersetup{linkcolor=blue,citecolor=blue,filecolor=cyan,urlcolor=blue}
\shorttitle{Surface Activity Contributions to Solar RVs}
\shortauthors{Anisha \& Rajaguru}
\graphicspath{{./}{figures/}}

\begin{document}


\title{Dynamics of photospheric magnetic flux distribution and variations in solar RVs -
a study using HARPS-N solar and SDO observations.}

\author[0000-0003-2694-3288]{ANISHA SEN} 
\email{anisha.sen@iiap.res.in}
\affiliation{Indian Institute of Astrophysics, II Block, Koramangala, Bengaluru 560 034, India}
\affiliation{Pondicherry University, R.V.Nagar, Kalapet, Puducherry 605014, India}

\author[0000-0003-0003-4561]{S.P. RAJAGURU}
\email{rajaguru@iiap.res.in}
\affiliation{Indian Institute of Astrophysics, II Block, Koramangala, Bengaluru 560 034, India}
\affiliation{Department of Physics and W.W. Hansen Experimental Physics Lab, Stanford University,
Stanford CA 94305, USA}
\altaffiliation{A major part of this work was completed when S.P.R was on sabbatical at Stanford University.}








\begin{abstract}
The distribution and evolution of photospheric magnetic field in sunspots, plages and network, and variations in their relative flux content, play key roles in radial velocity (RV) fluctuations observed
in Sun-as-a-star spectra. Differentiating and disentangling such magnetic contributions to RVs help in building models to account for stellar activity signals in high precision RV exoplanet searches. In this
work, as earlier authors, we employ high-resolution images of the solar magnetic field and continuum intensities from SDO/HMI to understand the activity contributions to RVs from HARPS-N solar observations. Using well observed physical relationships between strengths and fluxes of photospheric magnetic fields, we show that the strong fields (spots, plages and network) and the weak internetwork fields leave distinguishing features in their contributions to the RV variability. We also find that the fill-factors and average unsigned magnetic fluxes of different features correlate differently with the RVs and hence warrant care in employing either of them as a proxy for RV variations. In addition, we examine disk averaged UV intensities at 1600 \r{A} and 1700 \r{A} wavelength bands imaged by SDO/AIA and their performances as proxies for variations in different magnetic features. We find that the UV intensities provide a better measure of contributions of plage fields to RVs than the Ca II H-K emission indices, especially during high activity levels when the latter tend to saturate.


\end{abstract}

\keywords{Exoplanets --- Radial Velocities --- Sun Activity --- Detection of Faculae, Plages, Networks, Sunspots --- Data Analysis --- Techniques}


\section{Introduction} \label{sec:intro}

The radial velocity (RV) technique measures the Doppler shifts in the spectrum of a host star to derive 
its wobble motion caused by the orbital motion(s) of planet(s) around it.  Since its use in the first 
definitive exoplanet detection \citep{1995Natur.378..355M}, it has remained as a key tool in the 
discovery and characterization of exoplanetary systems. Major efforts are currently underway towards 
achieving extreme precision RV (EPRV) of centimeters per second needed to detect Earth-size planets around 
Sun-like stars \citep{2023AJ....165..151N}(see also NASA EPRV Working group final report \citet{2021arXiv210714291C}).
However, it is also well recognised that convective flows and magnetic activity in the 
photospheres of host stars cause RV fluctuations, known as astrophysical noise or jitter, much larger 
than the wobble signals caused by the orbital motion of a planet \citep{2011A&A...528A...4B,2023AJ....165...98L}.  
The RV contributions from stellar 
surface magnetic fields are intimately related to one of the fundamental effects in magnetohydrodynamics 
(MHD), viz. magnetic forces modifying the fluid motion. In stars with outer convection zones like the 
Sun, fluid motions at the photosphere are in the form of convective granules, which have the upward 
moving, hence spectrally blue-shifted, hotter less dense plasma occupying a much larger area than the 
downward moving, hence red-shifted, denser and cooler plasma at their boundaries \citep{1981A&A....96..345D}. 
As the flux or area coverage (fill-factor) of strong enough magnetic fields increases, the effects of 
magnetic suppression of plasma motions dominantly appear as a reduction in the convective blueshift of 
photospheric spectral lines due to the fact that most of the stellar light comes from the bright 
blue-shifted plasma \citep{2018ApJ...866...55C,2019Geosc...9..114C}.
In addition, the magnetic fields themselves, depending on their sizes and strengths, introduce their own 
thermal changes and hence in their brightnesses: dark spots and bright faculae, which can cluster to 
form bright plages, can differ in their relative contributions to the above suppression of convective 
blueshifts and hence to RVs \citep{2014ApJ...796..132D,haywood2016sun,2017A&A...597A..52M}. 
Further complications can arise from additional characteristic flows, such 
as large amplitude Evershed flows, that not so dark penumbrae of spots harbor. There is also the 
additional component of more uniformly distributed supergranular and intergranular magnetic networks, 
commonly known in the solar physics literature as network and internetwork, which can still interfere 
with convective motions in a manner that may depend on their size (flux) and strength distribution. 
These network magnetic elements, more uniformly distributed over the stellar disk, do not cause 
significant photometric modulation, except over the time-scales of solar cycle. Clearly, there is a 
complicated set of MHD processes at play resulting in delicate imbalances in the photometric and 
spectroscopic signatures of different magnetic structures and hence in their contributions to RVs \citep{2018arXiv180308708A}.

For distant stars, we cannot directly observe the surface phenomena and hence cannot observationally 
remove the above stellar surface contributions to RVs and so the RV method is severely limited by the 
effects of stellar magnetic activity. Clearly, retrieval of a planetary wobble signal in RVs require a 
good understanding and modelling of all the contributions from the stellar surface magnetic fields of 
differing strengths and fluxes. The Sun, the only star on which we can directly observe in a resolved 
manner the different magnetic structures, is an excellent test case that allows us to investigate the 
stellar RV fluctuations. It is exactly for this purpose, a solar feed to the High Accuracy 
Radial-velocity Planet Searcher for the Northern hemisphere (HARPS-N) instrument was designed for 
independent spectroscopic measurements of solar RVs \citep{2015ApJ...814L..21D}. Solar observations at 
HARPS-N began in July 2015 and the first three years of solar RVs derived from several hours of 
observations each day have been released and are publicly accessible via the Data and Analysis Centre 
for Exoplanets {\footnote{\textcolor{blue}{https://dace.unige.ch}}} hosted at the University of Geneva. 
Using the first release of RVs derived using the HARPS-N Data Reduction System (DRS) along with 
contemporaneous disk-resolved continuum intensity and magnetic field data from HMI/SDO and Total Solar 
Irradiance (TSI) data from SORCE Total Irradiance Monitor (TIM), \citet{milbourne2019harps} have shown 
that the HARPS-N solar RV fluctuations arise mainly due to the large and bright magnetic regions 
occupying areas larger than 60 $Mm^2$ and that the smaller structures do not significantly contribute. 
In addition, they also showed that the chromospheric Ca II H-K flux index log$(R'_{HK})$ and the optical 
light curves would provide effective proxies for RV variations in the plage-dominated stars but not in 
the case of the low-activity stars, where the plage and network filling factors are comparable.  Prior 
to that, using solar RVs derived from HARPS observations of sunlight scattered off the bright asteroid 
4/Vesta, \citet{haywood2016sun} found that the RV variations induced by solar activity were mainly due 
to the suppression of convective blueshift from magnetically active regions. They also found that the 
disc-averaged line-of-sight magnetic flux was a better proxy for the activity-driven RV variations than 
the FWHM and BIS of the cross-correlation profile and the Ca II H and K activity index. In their latest 
study \citet{haywood2022unsigned} concluded that the unsigned magnetic flux was an excellent proxy for 
RV variations.

The difficulties of differentiating and disentangling contributions of magnetic spots, plages/faculae 
and more uniformly distributed network structures to the RVs were further highlighted by 
\citet{milbourne2021estimating}, who concluded that more detailed information on the feature-specific 
filling factors are needed to fully characterize the host stars through the spectroscopic and brightness 
indices such as S-index and TSI. As alluded to earlier, the highly non-uniform distribution of magnetic 
fields, in strengths and sizes, accompanied by their differing thermal (brightness) signatures is behind 
the above difficulties. In the present analysis, we factor in some of the well studied physics behind 
the magnetic structuring of the solar atmosphere while extracting feature-specific filling factors from 
the high-resolution magnetic and intensity images of the Sun from the Helioseismic and Magnetic Imager 
(HMI) and Atmospheric Imaging Assembly (AIA) onboard the Solar Dynamics Observatory(SDO). We use the 
latest release of HARPS-N solar data calibrated using ESPRESSO DRS 2.3.5 (Section \ref{sec:data}) and 
examine more closely their correlated variations with different feature-specific fill-factors and 
average unsigned magnetic fluxes with a focus on assessing the contributions of intrinsically weak 
magnetic features, known in the solar physics literature as internetwork fields, which are known to 
contribute to brightness variations in certain wavelength bands \citep{2019LRSP...16....1B}. In general, 
we focus on the dynamical relationships between the flux distribution of different features, especially 
on the time-scales of decay of strong fields into weak fields, and their possible signatures in RV 
variations. We attempt to factor in established physics behind flux - strength 
relationships of surface magnetic structures to gain a better understanding of range and magnitude of RV 
fluctuations that can be expected in Sun-like stars with varying levels of interactions between 
convection and magnetic fields.

The paper is orgnaised as follows: Section \ref{sec:data} gives descriptions of the data used followed 
by, in Section \ref{sec:method}, of the methods adopted for the analysis, especially a detailed 
description of our new methods to identify the weak internetwork magnetic fields from HMI observations 
in Section \ref{sec:sub_method}. We present our results in Section \ref{sec:result} with subsections 
devoted to our new findings on (i) variations of, and connections between, feature-specific 
fill-factors and average unsigned magnetic fluxes, (ii) correlations between the SDO/HMI-derived 
magnetic quantities and RVs \& Ca II H-K flux indices log$(R'_{HK})$ derived from HARPS-N observations, (iii) 
a detailed analysis of new disk-averaged 1600 \r{A} and 1700 \r{A} UV intensities from 
SDO/AIA observations as magnetic proxies for RV variations, and (iv) on Lomb-Scargle periodogram analysis of 
time-scales of variations from magnetic features that are correlated with that in RVs, log$(R'_{HK})$s and also 
the SORCE Total Solar Irradiance (TSI) variations. Detailed discussions and our conclusions are presented in Section 
\ref{sec:conclusion}.

\begin{figure*}[!htbp]
\gridline{\fig{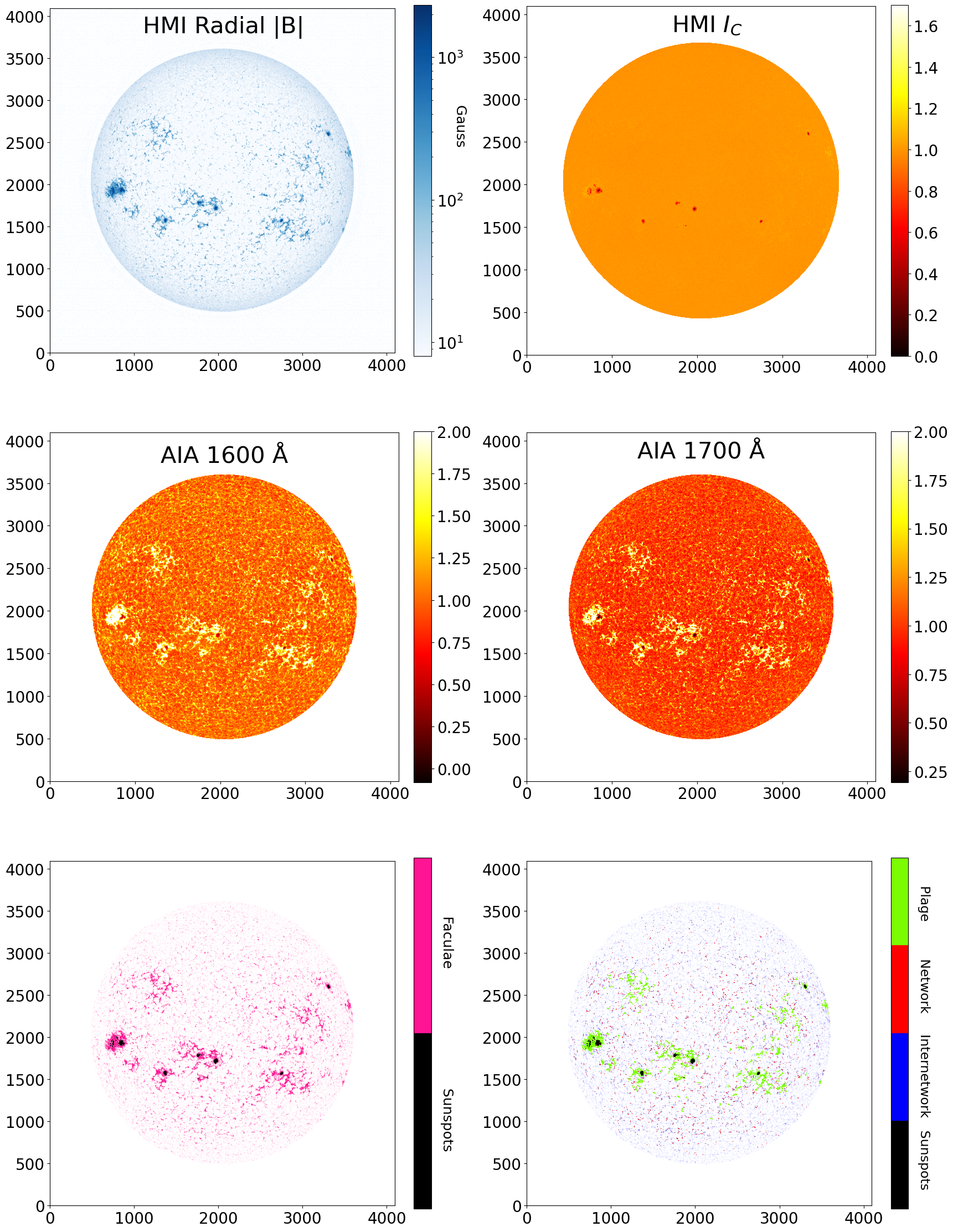}{0.9\textwidth}{}
          }
\caption{Sample images of SDO observables, from the 1st of January 2015:
radial magnetic fields from SDO/HMI LOS magnetogram (top left) and limb-darkening-corrected continuum intensity (top right),
SDO/AIA intensities at 1600 \r{A} (middle left) and at 1700 \r{A} (middle right). The color bar for HMI 
continuum intensity and AIA images saturate at a value well below the maximum value. 
Lower panels show thresholded images for the same date: the left one shows sunspots and all the 
rest of magnetic fields based on the thresholding criteria of \citet{haywood2016sun}, while the right panel shows 
the non-sunspot fields separated into plage, network and weak internetwork based on our new thresholding 
criteria explained in Section \ref{sec:sub_method}.}
\label{fig:1}
\end{figure*}

\section{Data} \label{sec:data}
\subsection{HARPS-N Solar Observations}
\label{subsec:harps}

The HARPS-N is a high-precision radial-velocity spectrograph installed at the Italian Telescopio 
Nazionale Galileo (TNG), a 3.58-meter telescope located at the Roque de Los Muchachos Observatory on the 
island of La Palma, Canary Islands, Spain.  Since July 18th, 2015, with its solar feed, HARPS-N has 
observed the Sun every day with a 5-minute cadence \citep{2015ApJ...814L..21D}. We have used these 
Sun-as-a-star spectroscopic data from the HARPS-N \citep{phillips2016astro, collier2019three}, as 
calibrated and released using the latest version of the pipeline ESPRESSO DRS 2.3.5 
\citep{dumusque2021three}. This release covers roughly a 3-year period between July 18, 2015 and 
December 31,2018. We use the pipeline-extracted RVs and the chromospheric flux index
$log(R'_{HK})$\citep{1984ApJ...279..763N}. The above data are publicly accessible at the Data and 
Analysis Centre for Exoplanets {\footnote{\textcolor{blue}{https://dace.unige.ch}}} hosted at the 
University of Geneva.

\subsection{SDO Observations}
\label{subsec:sdo}

Solar Dynamics Observatory (SDO) is a NASA spacecraft that has been observing the Sun since March 2010. 
Helioseismic and Magnetic Imager (HMI) \citep{scherrer2012helioseismic} onboard SDO observes the 
photosphere while the other two instruments, Atmospheric Imaging Assembly (AIA) 
\citep{lemen2011atmospheric} and Extreme Ultraviolet Variability Experiment (EVE) observe the 
chromospheric and coronal layers in various UV and E-UV wavelength bands. Data from this mission is 
publicly available {\footnote{\textcolor{blue}{http://jsoc.stanford.edu/}}}. For our analysis, we use 
HMI and AIA observations of the photosphere and chromosphere, respectively, for the same dates as 
HARPS-N Sun-as-a-star spectroscopic observations. HMI makes full-disk photospheric observations of 
continuum intensity, line-of-sight (Doppler) velocity and magnetic fields at 45 sec cadence, and also 
all the Stokes paramaters to derive the vector magnetic field at 135 sec cadence (although the standard 
data product is at 720 sec cadence). The above observables are derived from imaging over six wavelength 
positions across the Fe I 6173 \r{A} line at a spatial resolution of 1$''$ (pixel size of 0.5$''$) 
using a 4K $\times$ 4K CCD each for the LOS and vector quantities 
\citep{scherrer2012helioseismic,schou2012design,liu2012comparison,wachter2011image}. We have used HMI 
LOS magnetograms with a cadence of 45 seconds and limb-darkening-removed continuum intensities extracted 
at 720 seconds cadence. We use AIA full-disk UV intensities at the wavelengths 
1700 \r{A} and 1600\r{A} that image the upper photospheric and lower chromospheric layers with mean
formation heights of 360 km and 430 km above the photosphere \citep{2005Natur.435..919F}, 
respectively. AIA observes these wavelengths at a cadence of 24 seconds \citep{lemen2011atmospheric}.

The HMI LOS magnetograms and continuum intensities are used for identifying and 
separating different features on the solar surface and the UV intensities 
from AIA to estimate disk-averaged chromospheric emissions due to magnetic fields. We extract 24 
observations per day (one image per hour), spread over the duration between 1st January 2015 to 31st 
December 2018, from both the instruments (HMI and AIA), and average the derived quantities (as in 
Section \ref{sec:method}) for each day.

\subsection{SORCE/TIM TSI Observations}
\label{subsec:tsi}

Simultaneous high-accuracy Sun-as-a-star photometric observations are well known as key measures of 
solar activity variations from days, months to solar cycle time scales. Such information is also crucial 
to assess the relative contributions of different magnetic features, bright and dark, to the RV 
variations. For this purpose, and especially to ascertain further some of our newly identified 
contributions from weak inter-network fields, we employ Total Solar Irradiance (TSI) measurements by the 
Total Irradiance Monitor (TIM) onboard SORCE satellite \citep{2005SoPh..230...91K,2005SoPh..230..111K}. 
TSI data from SORCE/TIM are publicly available {\footnote{\textcolor{blue} 
{https://lasp.colorado.edu/home/sorce/data/tsi-data/}}}, and we employ a cotemporaneous 24-hour cadence 
time series of this data in our analysis.\\

\section{Method of Analysis} \label{sec:method}

A primary aim in this work is to factor in some well studied physics behind the flux and strength 
distributions of solar surface magnetic fields, especially on small spatial scales and flux contents, 
while identifying and estimating feature-specific fill-factors and their average unsigned magnetic 
fluxes thereby improving our understanding of their contributions to RVs. To this end, and also to 
compare with previous results, we follow the same basic steps in the preparation and processing of 
SDO/HMI full-disk images as originally done by \citet{haywood2016sun} and adopted in various follow up 
studeis ({\it e.g.,} \citet{milbourne2019harps,milbourne2021estimating,haywood2022unsigned}). 
We briefly describe the basic processing steps below, followed by a description 
of new features in our analysis in the next sub-Section.

We convert SDO images from pixel coordinates to heliographic coordinates (a coordinate system centered 
on the Sun) \citep{thompson2006coordinate} and employ a built-in routine {\it aiaprep} available in the 
{\it Sunpy} packages \citep{sunpy_community2020} to align the HMI and AIA images to a same spatial 
scale. We crop the HMI as well as the AIA images at a center-to-limb distance of 0.96 $R_{\sun}$ to 
avoid noisy pixels near the limb. For the AIA images, we employ a median filtering method 
\citep{lefebvre2005solar,bertello2010mount, chatterjee2016butterfly,bose2018variability} to remove limb 
darkening. Following \citet{haywood2016sun}, assuming that much of the magnetic flux on the solar 
surface is vertically oriented, we convert HMI LOS magnetic field strength $B_{obs}$ to radial magnetic 
field strength $B_{r}$ by removing the foreshortening effect, $B_{r,ij} = B_{obs,ij} /\mu_{ij}$, where 
$\mu_{ij} = cos \theta_{ij}$ and $\theta_{ij}$ is the angle between the outward normal on the solar 
surface and the direction of the line-of-sight from the SDO spacecraft. The 
upper left panel of the Fig.\ref{fig:1} shows an unsigned radial magnetogram, after correcting the 
foreshortening effects, and the upper right panel shows limb-darkening-corrected HMI continuum intensity and the 
middle panels show limb-darkening-corrected AIA 1600 \r{A} (left) and 1700 \r{A} (right) intensities. These 
images are from 1st January 2015. The noise level in HMI magnetograms is the lowest for pixels near the 
center of the CCD (around 5G) and increases towards the edges, reaching 8G at the solar limb 
\citep{yeo2013intensity}. Denoting the magnetic noise level in each pixel as $\sigma_{B_{obs,ij}}$,
we set $B_{obs,ij}$ and $B_{r,ij}$ to 0 for all pixels with a line-of-sight magnetic field measurement 
$B_{obs} < \sigma_{B_{obs,ij}} = 8 G $ \citep{haywood2016sun}.
\citet{yeo2013intensity} investigated the intensity contrast 
between the active and quiet photosphere using SDO/HMI full-disk images and found a cutoff at $|B_{r,ij} 
| > 3 \sigma_{B_{obs,ij}} /\mu_{ij}$. The separation of magnetic pixels from non-magnetic or quiet ones follows this 3$\sigma$ 
criterian, $ |B_{r,thresh1,ij} | = 24 G /\mu_{ij} $, and the further division of magnetic pixels into 
bright, called generally as faculae, and dark sunspots follows the same intensity thresholding crieria 
employed by \citet{yeo2013intensity} and \citet{haywood2016sun}:$ I_{thresh} = 0.89 I_{quiet}$, where $I_{quiet} = 
\frac{\sum_{ij} I_{flat,ij} W_{ij}} {\sum_{ij} W_{ij}}$ with weighting factor $W_{ij}$ = 1 if 
$|B_{r,ij}| < |B_{r,thresh1,ij}|$ and $W_{ij}$ = 0 otherwise. In the lower left panel of Figure 
\ref{fig:1}, we have shown a thresholded image separating sunspot (in black) from all the other fields 
(in purple). In the next step, we apply area thresholding to split the non-spot 
fields into plages and network: contiguous field patches exceeding an area threshold of 20 
$\mu$-Hemispheres or 60 $Mm^{2}$, the same as the one employed by \citet{milbourne2019harps}, are 
identified as plages (green patches in the lower right panel of Figure \ref{fig:1}).

\subsection{Identification of Weak Internetwork Magnetic Fields in HMI Observations}
\label{sec:sub_method}

The above processing steps separate flux outside of sunspots 
into bright (I $>I_{thresh}$) plages (area $>$ 60 $Mm^{2}$) and all the rest as network
\citep{milbourne2019harps,milbourne2021estimating,haywood2022unsigned}. However, it is well known that 
the quiet-Sun magnetic field on the solar surface has two fundamentally different distributions, in 
terms of their intrinsic strength and flux, due to their interaction with supergranular convection 
(\citet{hlin1995,solankietal96}, see \citet{2019LRSP...16....1B} for a detailed review): the quiet-Sun 
fields in the cell interior of supergranules, called as internetwork (IN, hereafter), are weak with 
typically less than or equal to the photospheric equipartition (kinetic) strengths of about 400 - 500 G 
with flux content in the range of $10^{16} -$ a few times $10^{17}$ Mx, while the network (NE, 
hereafter) fields confined to the boundaries of supergranules are made up of flux elements that have 
undergone ``convective collapse" \citep{parker1978hydraulic,spruit1979convective} attaining 
super-equipartition kilo-Gauss strengths with flux content typically larger than about a few time 
$10^{17}$ Mx. A detailed observational study \citep{solankietal96} of field strength versus flux 
relationship of small-scale fields gives a rough flux limit of $\sim$ 3 $\times$ $10^{17}$ Mx that 
separates the collapsed kG fields (NE) from partially collapsed intermediate strength or weaker IN 
fields. Such an organisation of small-scale magnetic flux has also sound theoretical basis, which 
derives from the effects of radiative smoothing on the convective collapse mechanism 
\citep{venkat1986,rajaguru2000}. Recent very high-resolution and high polarimteric sensitivity 
observations \citep{2014ApJ...797...49G,2020A&A...644A..86P,2021A&A...647A.182C} confirm the above basic 
characteristics of NE and IN magnetic fields in the solar photosphere, and also give a resolved picture 
of a typical weak IN field: it is a low lying small loop with its highly inclined (linear polarisation 
causing) segment, over a granule, flanked by vertical (circular polarisation causing) field within 
intergranular lane. A typical NE field element is a vertical structure rooted in the supergranular 
boundary and extending high into the chromospheric layers. 

At the HMI resolution of 1\arcsec, IN fields with flux limit of 3 x $10^{17}$ Mx will present themselves 
with strengths up to 56 G ($= 3 \times 10^{17}$ Mx$/(7.3 \times 10^{7}$cm$)^{2}$). Although the early 
studies that established the presence and properties of IN fields were done at 1 - 2$''$ resolution 
\citep{1996ApJ...460.1019L,2002ApJ...573..431L}, current understanding gained from high-resolution 
observations \citep{2019LRSP...16....1B} shows that observations start to resolve the IN fields at 1$''$ 
resolution, that there exists much more IN flux at still smaller scales and that the average filling 
factors increase as resolution increases. This aspect of IN fields is now well established, especially 
after the Hinode space mission enabled detailed measurements of quiet-Sun magnetic flux 
\citep{2008ApJ...672.1237L}, which was also shown to require a local dynamo distinct from the one 
generating the active region flux. 
Our results here (presented in the following Sections) show that, despite its lower resolution of 1$''$, 
HMI does capture a significant amount of IN flux. 
And as we show in later Sections, we are able to distinguish the differing signatures of IN 
and NE fields in the HARPS-N RV variations.

We note that the basic analysis step of \citet{haywood2016sun} converting the HMI LOS magnetic field 
strength $B_{obs}$ to radial magnetic field strength $B_{r}$ assumes that the most of the flux in the 
HMI magnetograms are vertically oriented. This assumption is still reasonable as high-resolution 
observations discussed above do indeed show circular polarization signals arising from the vertically 
oriented legs of IN fields, although it is expected that the $\mu_{ij} = cos \theta_{ij}$ correction may 
over-correct the contributions from the horizontal parts of IN field loops, which align to LOS as 
center-to-limb distance increases. However, such a systematics is taken care as we do include the 
$\mu_{ij}$ factor to the threshold of 56 G estimated from the above discussed flux limit of $\sim 3 
\times 10^{17}$ Mx that separates the weak and strong (super-equipartition) fields. 
Thus, in our final step, we split the non-spot (I $>I_{thresh}$ and B 
$>|B_{r,thresh1,ij}|$ ) and non-plage magnetic ( Area $< 60 Mm^2$ ) pixels into NE and IN fields 
depending on their magnetic flux density: those greater than the threshold value, $|B_{r,thresh2,ij}| = 
56 G /\mu_{ij} $, are grouped as stronger NE fields and the rest as weak IN fields. 
In the lower right panel of Fig.\ref{fig:1}, we have shown the result of above segmentation criteria that result in separated spot, 
plage, NE and the weak IN regions.

\section{Results} \label{sec:result}

\subsection{Feature-Specific Fill-factors and Average Unsigned Magnetic Fluxes}
\label{subsec:f_and_fB}

\begin{figure}[!htbp]
        \gridline{\fig{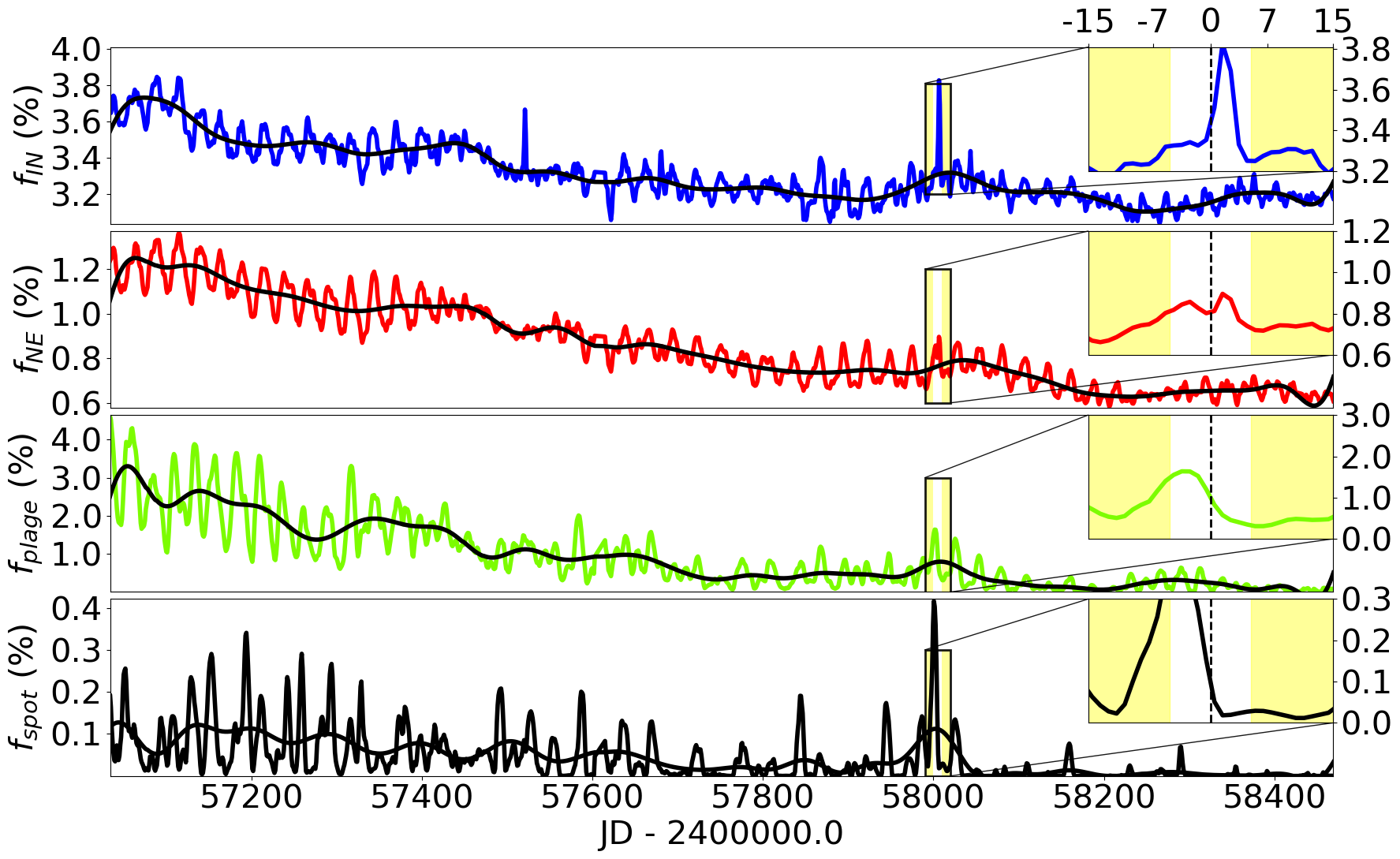}{0.48\textwidth}{}
          }
\caption{Feature-specific fill-factors, $f_{m}$, derived from full-disk SDO/HMI observations (top to bottom):
	weak internetwork (IN), strong network (NE), plage and sunspot fields. The overplotted black curve in each 
	panel is a smoothed one retaining only variations longer than 60 days obtained with a low pass Fourier filter.
	The insets in each panel show a zoomed-in view of a selected 30-day window around a large increase in spot flux 
	and its connections to other features.}
\label{fig:2}
\end{figure}

\begin{figure}[!htbp]
        \gridline{\fig{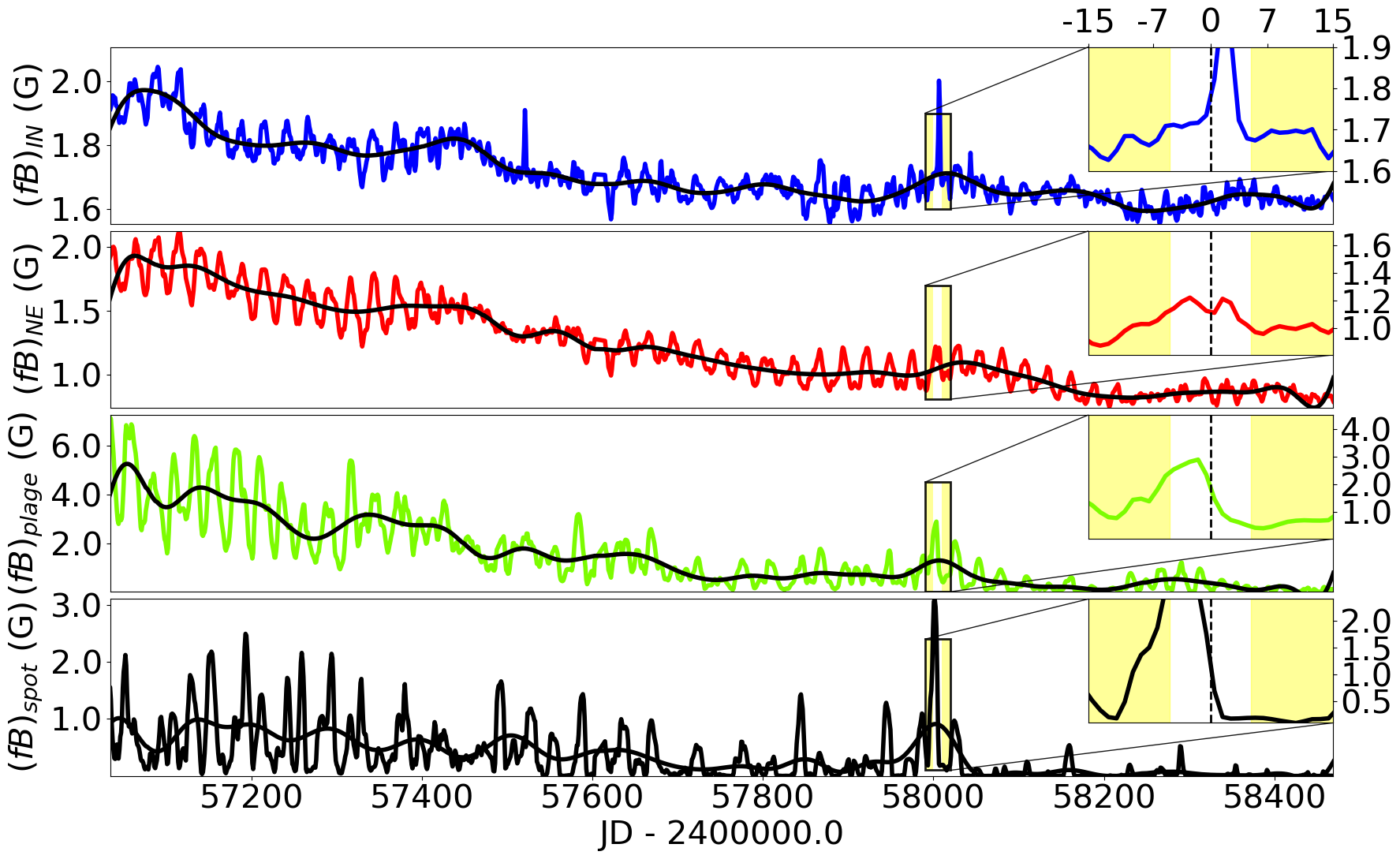}{0.48\textwidth}{}
          }
\caption{Feature-specific disk averaged unsigned field magnetic fluxes, $(fB)_{m} = f_{m}<|B_{m}|>$,
in the same (top to bottom) order, with the overplotted balck curves obtained in the same way, as in Figure \ref{fig:2}.
	The insets are the same as explained in the caption of Figure \ref{fig:2}.}
\label{fig:3}
\end{figure}

A basic feature in our analysis, distinct from previous ones, of HMI magnetic field observations is the 
identification and separation of the IN fields, as explained in the previous Section. Figure \ref{fig:2} 
shows the fill factors of IN, NE, plage and spot fields for the full 4-year period (1437 days) between 
January 2015 to December 2018. Note that the HARPS-N observations (shown in later Figures) cover a 
subset of days, totalling about 609 days, within this period (due to missing days and a longer gap towards the end).
We have also calculated feature-specific average unsigned magnetic fluxes, $<|B_{m}|> = 
\frac{\sum_{ij}|B_{m,ij}|}{N_{m}}$, where the subscript $m$ stands for spot, plage, NE or IN, $B_{m,ij}$ 
is the pixel field strength and $N_{m}$ is the total number of pixels occupied by feature $m$. The disk 
averaged unsigned magnetic flux of the different features are then $f_{m}<|B_{m}|>$, which we simply 
write as $(fB)_{m}$ and plot them in Figure \ref{fig:3}. The total disk-averaged unsigned magnetic flux 
is then $<|B|> = \sum_{m}f_{m}<|B_{m}|>$, which is simply the sum of the different panels of Figure 
\ref{fig:3}. The starting date (January 2015) falls during moderately high activity -- about a year 
after the maximum of Solar Cycle 24, and ending date in December 2018 is around the cycle minimum, and 
hence all the quantities, including the RVs and $log(R'_{HK})$ from HARPS-N, show a clear decline over 
the analysis period. 
Variation at solar rotation period is dominant in the feature specific fill-factors $f_{m}$ and average 
unsigned magnetic fluxes $(fB)_{m}$ shown in Figures \ref{fig:2} and \ref{fig:3} -- the overplotted
smooth curves in black in the different panels correspond to longer time-scale variation obtained
after a Fourier filter (low-pass filter) to remove periods shorter than 60 days. 

We note several interesting and dominant patterns along with subtle but still easily discernable 
differences and similarities between the features: (i) the relative change in IN fields, over the longer 
cyclic time-scale, is an order of magnitude less compared to those of other features, (ii) the IN fields 
show a shorter time-scale noise-like fluctuations, (iii) within the intermediate time-scales of longer 
than rotation and upto a year, the IN and NE fields show a correlated variation on the longer period 
side while the plage and spot fields correlate tightly on time scales of a few months, (iv) spots show 
larger fluctuations on the rotation as well as slightly shorter and longer time-scales, and (v) there 
are intermittent instances highly correlated variation, with a few days to a week of time lag, between 
spots and the IN fields (see the insets in Figures \ref{fig:2} and \ref{fig:3}) -- the two large peaks 
in $f_{IN}$ and $(fB)_{IN}$ are associated with a similar increase in $f_{spot}$ and $(fB)_{spot}$ about 
a week earlier --, while there is no such correlation between spots and NE at these two instances. The 
above noted features carry important information on the dynamical connections between the evolution of 
these different magnetic flux concentrations and their interactions with convective and other 
large-scale flows on the Sun. In the context of RV variations, which we address in the Sections to 
follow, we particularly note the short time-scale fluctuations of IN fields, which while at the same 
time change relatively little over the longer solar cycle time-scale compared to other features. This 
latter aspect of IN fields is also well appreciated in the solar literature \citep{2019LRSP...16....1B}. 
We discuss further these features in Section \ref{subsec:periodogram}, where we analyse the time-scales involved 
through periodograms.

To understand further and check the new features of our analysis, we compare our estimates of 
$f_{m}$ in Figure \ref{fig:2} with those of \citet{milbourne2021estimating,haywood2022unsigned}, especially for the network flux, 
which in our analysis have been separated into weak IN and strong NE. We find that the sum $f_{IN}+f_{NE}$ of IN and NE fill-factors 
(sum of top two panels of Figure \ref{fig:2}) is more than twice the estimate for $f_{ntwk}$ in Figure 2 of \citet{milbourne2021estimating}
(or Figure 4 of \citet{haywood2022unsigned}) while the plage ($f_{plage}$) and spot ($f_{spot}$) fill-factors match closely.
Since this is a rather large discrepancy, we set out to examine all the differences in the way the data were processed 
and analysed. We find that our results and that of \citep{haywood2016sun} match exactly for fill-factors of spot and all 
non-spot fields, which are termed as faculae by \citet{haywood2016sun}, while all the later studies published by 
\citet{milbourne2019harps,milbourne2021estimating,haywood2022unsigned} show the above discrepancy that we have noted for the network 
fields (IN and NE). A careful examination reveals that all these later studies have employed
720 sec cadence data, while ours here and of \citet{haywood2016sun} use the original 45-sec cadence data
from HMI. We clarify that the 720 sec HMI data are actually averages of 45 sec cadence basic observations over 720 seconds
\citep{hoeksema2014helioseismic}. Hence, noting that internetwork (IN) fields are typically moved around by granules
with typical life times of 5 - 10 minutes, it is expected that the 12-minute (720 sec) averaging of HMI observations will
smooth out the IN fields due to granular time-scale displacements as well as due to cancellations of opposite polarity signals
passing through a given location over this time interval. Hence, use of 720 sec exposure data from HMI will yield
significantly reduced values for the fill-factors of IN fields ($f_{IN}$), and this certainly plays a role in the much
reduced values for $f_{ntwk}$ of \citet{milbourne2021estimating}.
To test this explicitly, we have repeated our analysis using the 720-sec exposure data from HMI and the results and comparisons
are presented in Appendix A in a table and in a figure. In summary,
we now find that the total network flux (IN + NE flux), $f_{IN}+f_{NE}$, that we find in our main analysis (in Figure \ref{fig:2}),
is much larger than those reported by \citet{milbourne2021estimating,haywood2022unsigned} mainly because of the latter authors
using HMI magnetograms averged over 12 minutes, which misses much of the weak IN fields evolving over granular convection time scales.
Further, as we show in Appendix A, fill-factors ($f_{NE}$) of stronger (or collapsed) NE fields do not show any difference between
the use of 45 sec or 720 sec cadence HMI data, because the NE fields typically have a much longer lifetime (20 minutes
or longer) than the IN fields \citep{2019LRSP...16....1B}. This reaffirms that our method to separate the weak (IN) and strong (NE) 
network fields based on a flux criterion has indeed worked successfully.
We conclude that our identification criteria respecting the physics behind the dynamics 
of small-scale magnetic fields along with our use of original 45 sec cadence and 1$''$ resolution HMI data, 
compared to the 12-minute averaged and a likely lower effective resolution (due to 2x2 binning) employed by \citet{
milbourne2019harps,milbourne2021estimating,haywood2022unsigned}, have facilitated capturing especially the 
weak IN fields, which are missed in these earlier measurements. This weak component of network fields, as we show
below, carries a large fraction of total solar magnetic flux but changes slowly over the solar cycle time scale and hence is
important to determine the base level of RV fluctuations due to magnetism.
Note that differences in temporal and spatial resolutions 
do not affect measurements of plages and spots and hence we match earlier measurements for these larger features.

\begin{figure}[!htbp]
\gridline{\fig{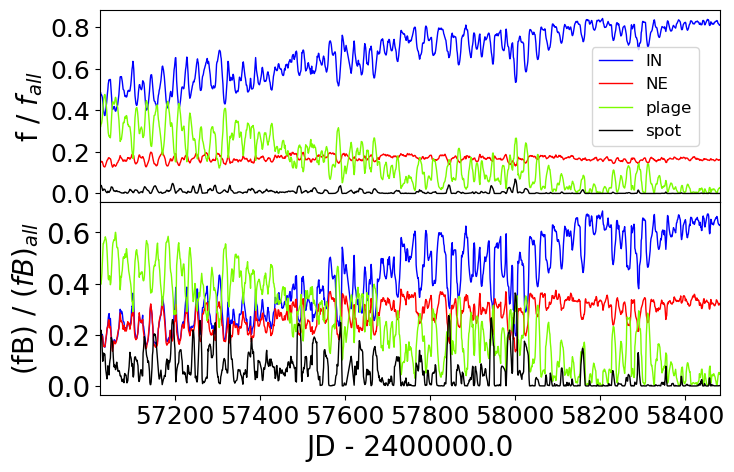}{0.47\textwidth}{}
          }
\caption{Time evolution of relative areas, $f/f_{all}$, and average unsigned magnetic fluxes, $(fB)/(fB)_{all}$, over the
	4-year period January 2015 to December 2018.}
\label{fig:4}
\end{figure}

To elucidate further the connections between different magnetic features, we plot in Figure \ref{fig:4}
relative contributions of individual $f_{m}$ and $(fB)_{m}$ to the total for the whole Sun. Firstly, results here 
indeed show that there is a siginificant fraction of solar magnetic flux in the IN as evidenced in the 
bottom panel of Figure \ref{fig:4}: at cycle maximum about 20 - 25\% of solar magnetic flux is in the weak form, 
which increases to more than 50\% near cycle minimum. In terms of numbers for the fluxes, 
we find that, within the time-period of observations covered, the IN and NE fields have similar amount of flux, 
$0.8 - 1.2\times 10^{23}$ Mx  and $0.5 - 1.3\times 10^{23}$ Mx, respectively (see the right panel of Figure \ref{fig:5}). 
High-resolution observations by \citet{2013SoPh..283..273Z,2014ApJ...797...49G} estimate that IN fields carry a total 
flux of the order of $\sim 1.1 \times 10^{23}$ Mx and NE fields at a much higher value of $\sim 6.8 \times 10^{23}$. 
This shows that HMI, at 1$''$ resolution, misses a significant amount of network fields (IN and NE fields), especially 
much of the very small-scale collapsed kG strength NE fields. It is also possible that the flux limit of 
$3 \times 10^{17}$ Mx and hence the HMI flux density of 56 G that
we used to separate IN from NE fields is not accurate, implying there still exist kG fields (i.e., NE) at lower flux contents. 
In this situation, our identified IN and NE fields may have intermixed contributions. In any case, HMI measurements 
yielding a lower flux for IN and NE fields together is expected and it is certainly due to the lower resolution and 
thereby missing some scale-scale weak as well as strong flux. For active regions (spots + plages), our estimated flux is 
$\sim 6.5 \times 10^{23}$ Mx at the maximum activity level within the time period covered in Figure \ref{fig:5}.
This compares well with similar estimates ($\sim 6 - 8\times 10^{23}$) for cycle maximum \citep{1994SoPh..150....1S,
2014ApJ...797...49G}, noting that the maximum activity level we have in our data is about 1 year past the Cycle 24 maximum.

\subsubsection{Correlations between Magnetic Fluxes, Fill-factors and Strengths}
\label{subsubsec:f_and_B}
\begin{figure*}[!htbp]
\gridline{\fig{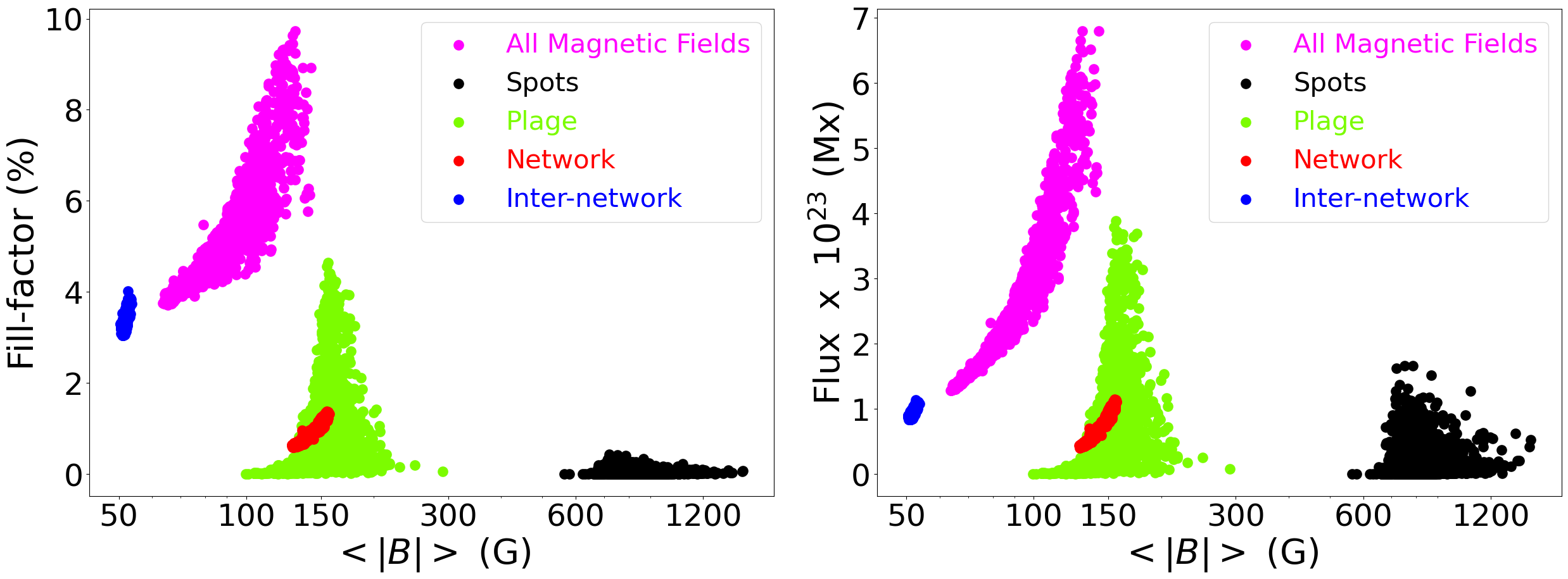}{0.9\textwidth}{}}
\caption{Relation between area fill-factors ($f_{m}$, left panel) or total magnetic fluxes (right panel) of different magnetic 
	features and their disk-averaged unsigned field strengths ($|B_{m}|$).}
\label{fig:5}
\end{figure*}
The dependence of intrinsic field strength of a solar magnetic structure on its flux content is a
key relationship that derives from the physics of magnetic field intensification in the near-surface layers
\citep{parker1978hydraulic,solankietal96,venkat1986,rajaguru2000}
and we used that to separate the weak IN fields as discussed in Section \ref{sec:sub_method}.
Despite carrying a large amount of flux (cf. Figure \ref{fig:3}), since they are in a shredded weak form 
at sub-granular scales, the IN fields interfere with convection in a different manner than the
collapsed strong fields comprising plages and spots: for these latter structures increasing flux primarily
increases their areas (fill-factor) replacing the convective granules and thus directly contributing to
reduction of convective blueshift (and thus to RV variations). The transition between weak fields to strong fields
is not sharp, but through a characteristic relationship between flux and strengths \citep{solankietal96,
venkat1986,rajaguru2000}, which we indeed see exhibited by the network fields: 
Figure \ref{fig:5} shows the relationship between the unsigned average field strengths ($|B_{m}|$) 
and fill-factors ($f_{m}$) (left panel) and total fluxes ($(fB)_{m}\times Area_{\odot}$, right panel) for the 
different features along with that for the whole of magnetic fields on the Sun. Firstly, we note that our flux-per-feature 
criterion ($3 \times 10^{17}$ Mx or $56$ G pixel strength at HMI resolution) has indeed very well separated the weak IN field from the 
strong NE fields, which have the same strengths as plage fields and differ only in their sizes (flux contents). 
Secondly, as expected, the average unsigned field strengths of spots and plages show no trends against 
fill-factors (areas) or their flux contributions, whereas that of network fields (IN and NE) show a strong correlation.
The above flux-strength relation for the whole of the magnetic field is also plotted in Figure \ref{fig:5} 
(data points in pink color), and it is clear that the exponentially increasing flux contributions over higher end of
$|B| =$100 - 150 G are mainly from collapsed strong fields of NE, plages and spots. This relationship is essentially
the same one as that studied by \citep{solankietal96} in their high-resolution observations of weak IN and the stronger
partially or fully collapsed NE field elements, and we have here verified it in HMI data.
These correlations between fill-factors $f_{m}$ and average
unsigned magnetic fluxes $(fB)_{m}$ have to be taken into account while assessing any correlations between
$f_{m}$ or $(fB)_{m}$ and the RV variations, $\Delta$RV, presented in the next Section. 

\begin{figure}[!htbp]
\gridline{\fig{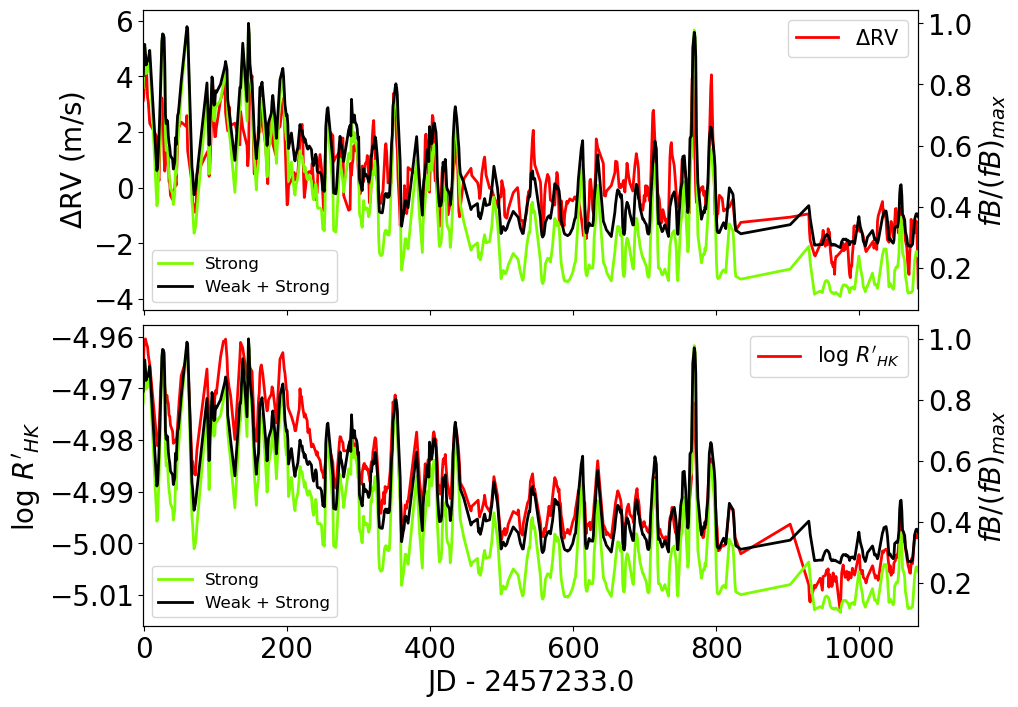}{0.5\textwidth}{}
          }
\caption{Comparison of variations in HARPS-N $\Delta$RV (upper panel) and in
$log(R'_{HK})$ (lower panel), in red, with those in average unsigned magnetic fluxes of the strong field 
($(fB)_{spot}+(fB)_{plage}+(fB)_{NE}$) (in green), and that including the weak IN field (in black) 
covering 609 days over 2015 - 2018. Note that $(fB)$s are normalised with their maximum values to enable a comparison
of the ralative variations.} 
\label{fig:6}
\end{figure}

\subsection{Correlations between HARPS-N RVs and SDO/HMI Magnetic Field Observations}
\label{subsec:corrs}
From the time series of RVs from HARPS-N, which has covered about a 609-days period between 2015 July 29 and 2018
July 16, we compare in Figure \ref{fig:6} the variations of mean-subtracted $\Delta$RV ($RV - <RV>$) and $log(R'_{HK})$ 
with those of feature-specific average unsigned magnetic fluxes $(fB)_{m}$ estimated from HMI/SDO observations. 
For this comparison, we use the sum of all $(fB)_{m}$ with and without the IN field ($(fB)_{IN}$) and label them
as ($strong + weak$) and ($strong$) fields, respectively; for a meaningful comparison here, since it is of 
different physical quantities and we need only their relative variations, we normalise a $(fB)_{m}$ by its 
maximum value. While very good correlation at rotation time-scales between these quantities is obvious, we note
that, more importantly, on longer time scales the inclusion of IN magnetic fields (black curve) matches the 
variations in $\Delta$RV and $log(R'_{HK})$ much more closely than without them (green curve). 
Such a contribution from IN fields is expected because flux contained in them is significant (cf.
Figure \ref{fig:3}) and moreover they dominate the total flux during the solar minimum period as evident in the relative 
variations of $(fB)_{m}$ shown in Figure \ref{fig:4}. Thus the IN fields change relatively little over the solar cycle
amounting to a nearly constant background flux on the Sun and cause a long-term background signal in $\Delta$RV.
However, we note that the true level of this background IN flux and hence the minimum or base level of 
variations in mean-subracted RVs do require covering fully a solar cycle minimum period. Otherwise, any accounting of
relative variations in $\Delta$RV that are biased by larger variations of active region fluxes would suffer from a long-term
offset as seen in Figure \ref{fig:6}.
It is also clear (from the bottom panel of Figure \ref{fig:6}) that the correlation between $log(R'_{HK})$ and the 
full magnetic flux (strong + weak fields) is much cleaner and tighter than that between $\Delta$RV and the magnetic flux. 
This is largely due to the shorter time-scale noise-like fluctuations in $\Delta$RV.

\begin{figure}[!htbp]
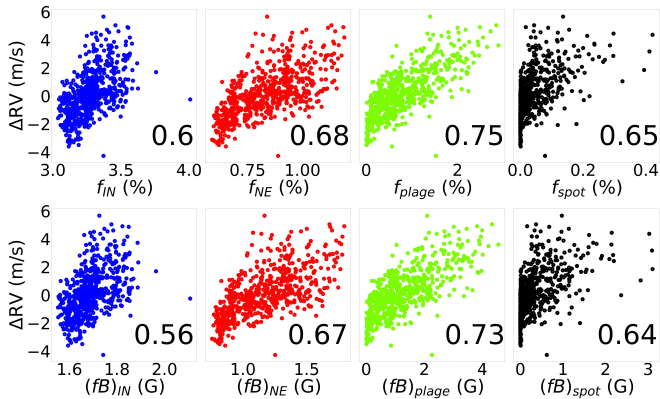

\gridline{\fig{fig7.pdf}{0.48\textwidth}{}
          }
	\caption{ Correlations between $\Delta$RV and the fill factors ($f_{m}$, upper panel), and average
	unsigned magnetic flux ($(fB)_{m}$, lower panel) of the different magnetic features, as indicated in axis labels.
	The Spearman correlation coefficients are given in each panel.}
\label{fig:7}
\end{figure}

\begin{figure}
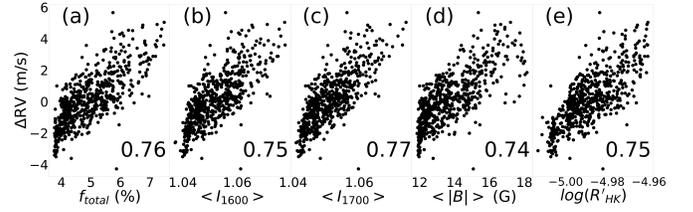

\gridline{\fig{fig8.pdf}{0.48\textwidth}{}
          }
\caption{Correlations of HARPS-N $\Delta$RVs with the five main parameters characterizing solar
magnetic contributions: total fill factor, disk-averaged UV intensites at 1600 \r{A} and 1700 \r{A}, average unsigned magnetic field,
and chromospheric Ca II-K flux index (left to right). Spearman correlation coefficients are given in each panel.}

\label{fig:8}
\end{figure}

Next, we examine scatter plots of $\Delta$RV ($RV - <RV>$) and the feature-specific $f_{m}$ and $(fB)_{m}$ and
their correlation coefficients: the top row of Figure \ref{fig:7} shows $\Delta$RV against $f_{m}$ while the bottom 
row is that between $\Delta$RV and the unsigned average magnetic flux $(fB)_{m}$. For each plot, we
computed the Spearman correlation coefficients to measure the degree of correlation between two variables. 
Plage fields, in conformity with previous results
\citep{milbourne2019harps}, show the largest correlation ($\sim 0.75$) with the variations in $\Delta$RV. 
Importantly, variations in IN fields show a significant correlation ($\sim 0.6$) with that in $\Delta$RV.
We note that the $\Delta$RV are, in general, better correlated with fill factors $f_{m}$ than with the average unsigned 
fluxes $(fB)_{m}$. Further, the reduction in the correlation between $\Delta$RV and $(fB)_{m}$ compared to that between
$\Delta$RV and $f_{m}$ is the largest for the weak IN fields.
As we pointed out earlier, differences between correlations of $\Delta$RV with $f_{m}$ and $(fB)_{m}$ could be influenced
by the relations between field strengths and fluxes shown (Figure \ref{fig:5}) and discussed in Section 
\ref{subsubsec:f_and_B} -- the relative changes in average field strengths (flux densities) for a given amount of 
change in fill-factor are much smaller for plages and spots than for network fields.
Correlations between $\Delta$RV and the total magnetic fill-factor $f (=\sum_{m}f_{m})$, 
full-disk-averaged unsigned magnetic flux $<|B|>$, UV intensities at 1600 and 1700 \AA~ (see next Section), 
and $log(R'_{HK})$ are shown in Figure \ref{fig:8}. Although the correlation coefficients (marked within the panels in 
Figure \ref{fig:8}) between the Sun-as-a-star spectroscopic quantities ($\Delta$RV and $log(R'_{HK})$ from HARPS-N)
and the disk-averages of resolved observations from SDO (HMI and AIA) all are very similar, we do note that 
the fill-factors correlate stronger than the unsigned average magnetic flux. Among the chromospheric quantities, we find the
disk-averaged UV intensities at 1700 \AA~ correlate the strongest with the $\Delta$RVs -- we discuss these 
chromospheric activity proxies in detail in the following Section.

\begin{figure}
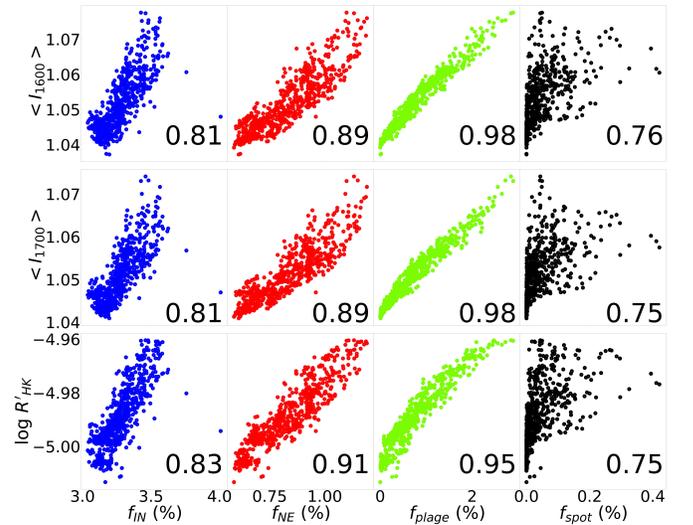

\gridline{\fig{fig9.pdf}{0.48\textwidth}{}
          }
\caption{Correlations of fill factors of the different magnetic features with the average UV intensities 
at the wavelengths 1600 \r{A} and 1700 \r{A}, and the chromospheric Ca II-K flux index (top to bottom). Spearman correlation 
coefficients are given in each panel.}

\label{fig:9}
\end{figure}

\begin{figure}
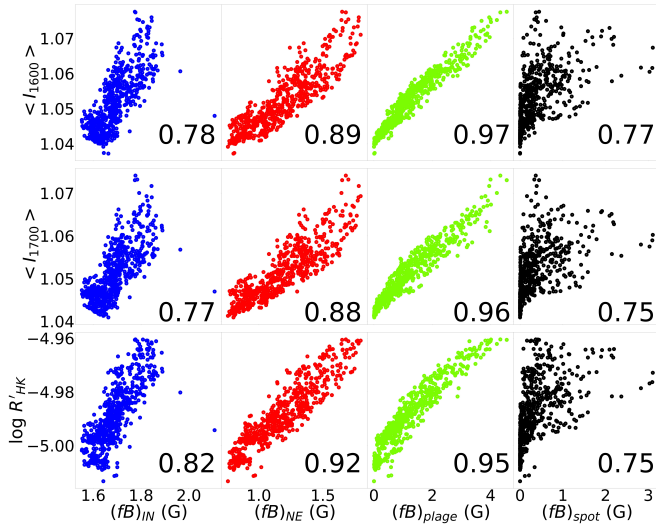

\gridline{\fig{fig10.pdf}{0.48\textwidth}{}
          }
\caption{Correlations of the average unsigned magnetic flux of the different magnetic features 
with the average UV intensities at the wavelengths 1600 \r{A} and 1700 \r{A}, and the
chromospheric flux index (top to bottom). Spearman correlation coefficients are given in each panel.}

\label{fig:10}
\end{figure}

\subsection{UV Intensities at 1600 \AA~ and 1700 \r{A} as Magnetic Flux Proxies}

The chromospheric Ca II K emission index $log(R'_{HK})$\citep{1984ApJ...279..763N} is a well known observational quantity
that acts as a proxy for the photospheric magnetic flux threading the chromospheric layers. 
It traces very well the strong supergranular network (NE) and the plage fields and hence correlates very well with the
RV fluctuations (see Figure \ref{fig:6}). It is also well known that UV emissions over wavelength bands
centered at 1600 \AA~ and 1700 \AA~ originating in upper photospheric and chromospheric layers faithfully capture
the underlying magnetic flux \citep{2001A&A...379.1052K}. Here, we experiment with the same
UV emission intensities imaged by SDO/AIA, extracted and processed as explained in Section \ref{sec:method}. The disk-averaged
intensities, $<I_{1600}>$ and $<I_{1700}>$, are derived from 1-hour cadence images averaged over a day (24 images) as for
other HMI observables that we employed. The scatter plot of $<I_{1600}>$ and $\Delta$RV is shown in
panel (b) and that between $<I_{1700}>$ and $\Delta$RV in panel (c) of Figure \ref{fig:8}; the correlation coefficients, respectively,
are 0.75 and 0.77. While the former is the same as that between $log(R'_{HK})$ and $\Delta$RV, the $<I_{1700}>$ correlate stronger
with $\Delta$RV. 
We also study the correlations of $<I_{1600}>$, $<I_{1700}>$ and $log(R'_{HK})$ with the feature-specific $f_{m}$ and $(fB)_{m}$
of IN, NE, plage and spot fields in Figures \ref{fig:9} and \ref{fig:10}. Comparing the correlation coefficients in these plots,
we find that UV intensities correlate significantly stronger with plages than the $log(R'_{HK})$.
A closer examination of panels for plage fields shows that the $log(R'_{HK})$ tend to saturate at the largest fill-factors
(or high activity levels), while the UV intensities remain more linearly correlated. Noting that the data period used
in this work ($i.e.,$ the period covered by HARPS-N Solar observations) starts well past the cycle 24 maximum,
we expect that the slight saturation that we see in Ca II K emission would likely be much larger at cycle maximum activity levels.
Hence, at high activity levels the UV intensities would provide a better measure of plage fields and hence
could be more reliable proxies for variations in RVs caused by these fields. We note here that the saturation of 
chromospheric Ca-II K emissions at high activity levels is well known and studied on the Sun (see, for example 
\citet{2009A&A...497..273L}) as well as in a large number of other stars (e.g. \citet{2022A&A...662A..41R}).

\begin{figure}[!htbp]
        \gridline{\fig{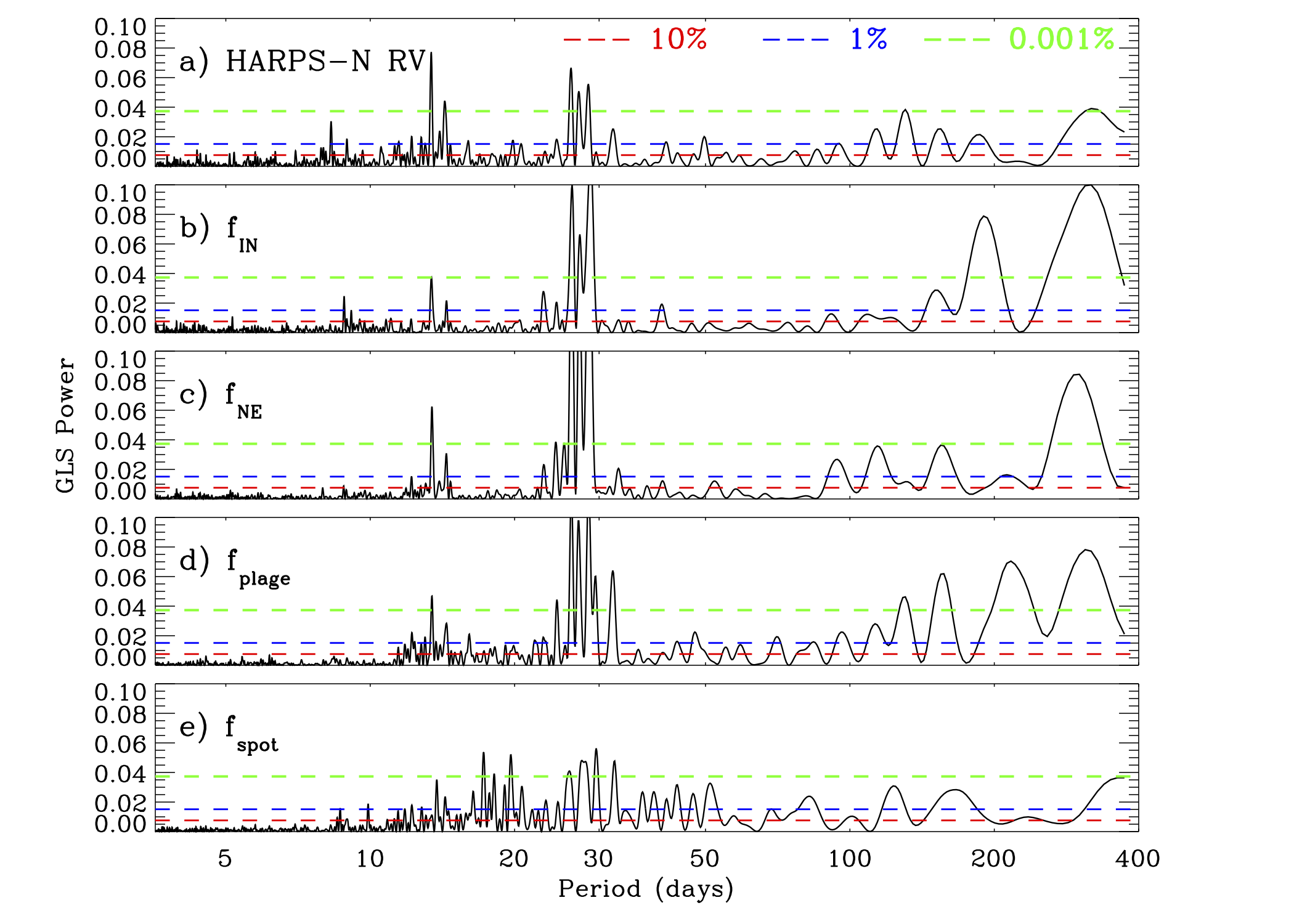}{0.54\textwidth}{}
          }
	\caption{Generalised Lomb-Scargle periodograms of feature-specific fill-factors $f_{m}$}
\label{fig:11}
\end{figure}

\subsection{Periodogram Analysis}
\label{subsec:periodogram}
\begin{figure}[!htbp]
        \gridline{\fig{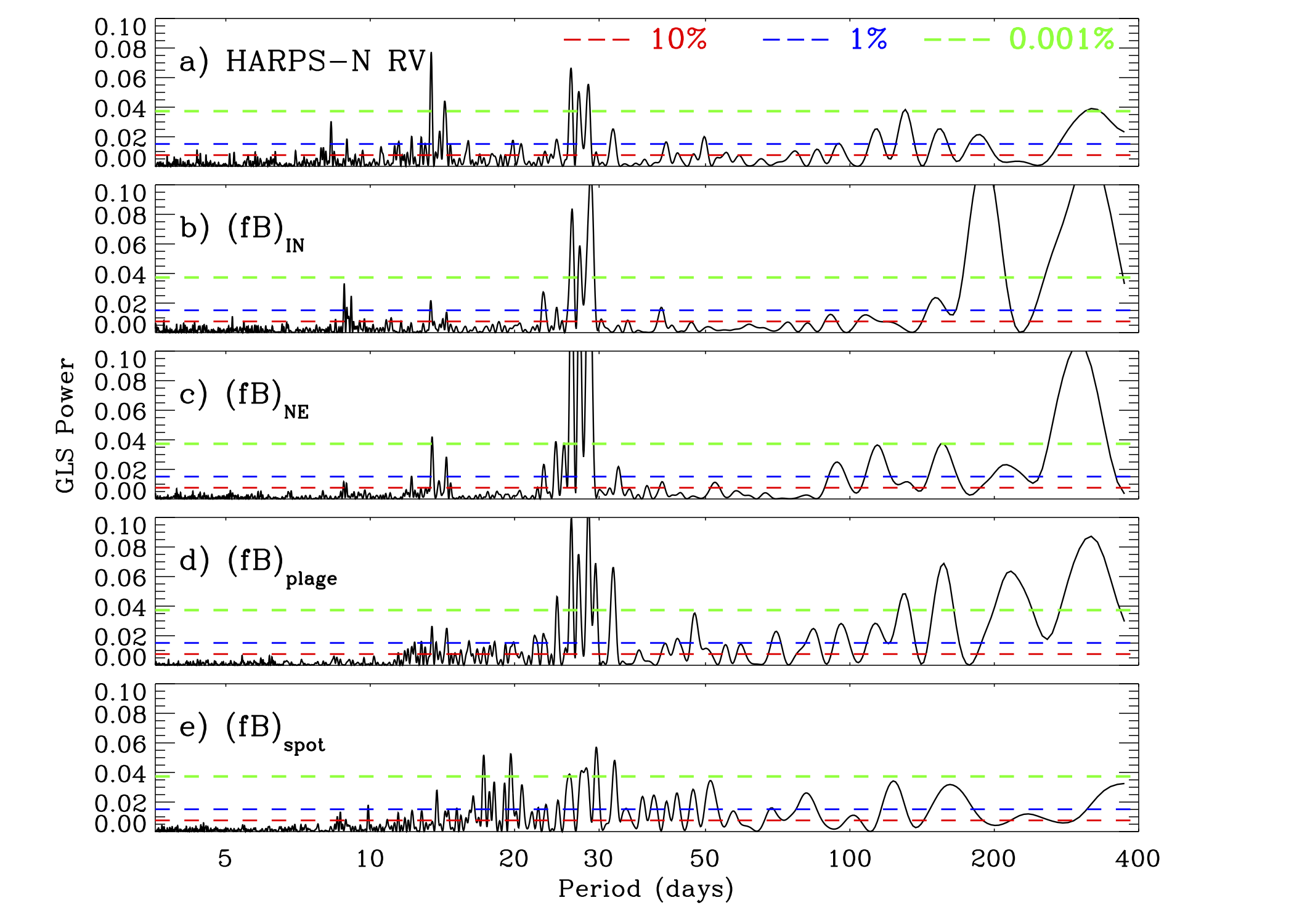}{0.54\textwidth}{}
          }
	\caption{Generalised Lomb-Scargle periodograms of feature-specific average unsigned magnetic fluxes, $(fB)_{m}$}
\label{fig:12}
\end{figure}

\begin{figure}[!htbp]
        \gridline{\fig{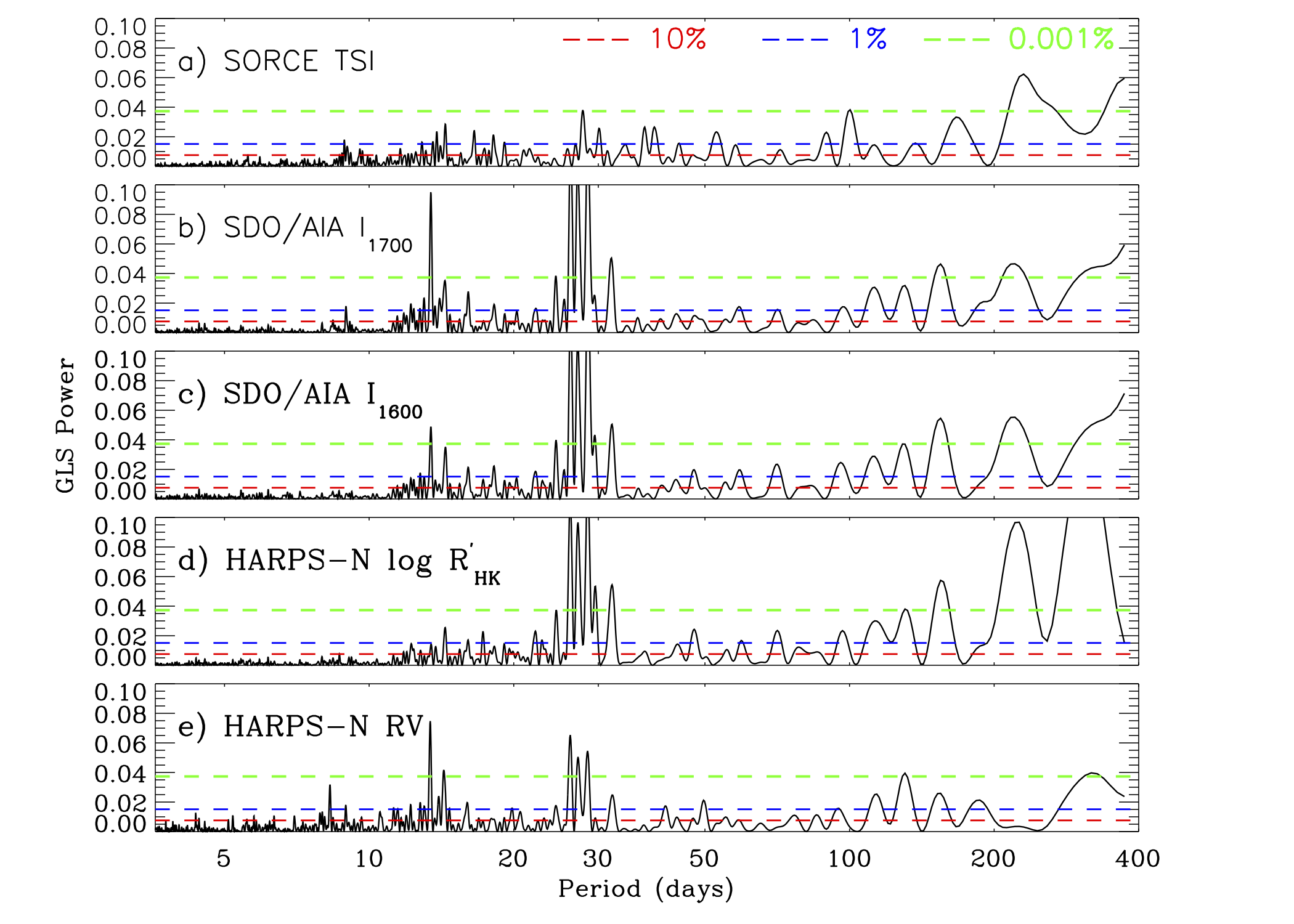}{0.54\textwidth}{}
          }
\caption{Comparison of GLS periodograms of photospheric TSI and $\Delta$RV with those of chromospheric 
	$log(R'_{HK})$ and $<I_{1600}>$.}
\label{fig:13}
\end{figure}

Lomb-Scargle periodograms are commonly used to detect periodic signals due to planets in the stellar RVs.
While presenting the time-series of $f_{m}$ and $(fB)_{m}$, in Section \ref{subsec:f_and_fB}, we already
discussed the typical time-scales introduced by the evolution of magnetic flux in the different features,
especially due to decay of active region flux into weak IN fields and longer time scales involved in the
reorganization of flux into NE fields. Here we examine their signatures that appear as power peaks in 
generalized Lomb-Scargle (GLS) periodograms \citep{2009A&A...496..577Z} and compare them with those of $\Delta$RV. 
The results are shown in Figures \ref{fig:11}, \ref{fig:12}, and \ref{fig:13}.
We mainly focus on shorter periodicities, especially the rotation period and shorter ones. Note also that
total time length of HARPS-N RVs is less than three years. Comparisons of GLS periodogram of $\Delta$RV with
that of feature-specific $f_{m}$s and $(fB)_{m}$s are shown in Figures \ref{fig:11} and \ref{fig:12}, respectively.
The dominant periodicities seen are of rotation (27 - 30 days) and its first harmonic (13 - 15 days) in all
the quantities, above the false alarm probability (FAP) of 0.001\%. We note, in particular, the very similar
periodicities present in the spectra of $\Delta$RV and the weak IN fields over the short period range 
between 7 and 10 days; although the FAP is between 10\% and 1\% for most of these short periods,
we note a peak significantly above 1\% FAP at 8 - 9 days for the IN fields in Figures \ref{fig:11} and \ref{fig:12}.
Although power peaks at 9 - 10 day periods would correspond to third harmonic of the 
primary rotation period (27 - 30 days), we note that the stronger NE and plage fields do not 
exhibit any significant peak at periods shorter than 10 days except perhaps the spot fields that have a dense set of
peaks over a wide range of periods. Hence we speculate that the 
origin of higher harmonics, especially the thrid harmonic period of 8 - 9 days in IN fields, is 
related to their latitudinal distribution and time-evolution, which cause perhaps some beating interference with the
rotation of sunspot belt. Such a possible physical origin of periods close to the third harmonic is strengthened further 
below when we compare the periodograms of photospheric and chromospheric quantities. 
Given that we capture only about one-third of the IN flux
that is present on the Sun from SDO/HMI observations, we speculate that much of the shorter periods present
in $\Delta$RV are likely due to the weak IN fields. The sunspot fill-factors and magnetic fluxes 
show a prominent cluster of peaks around 20 days, which are missing in the spectra of all the other quantities.
Such peaks have been noted by several authors, and their origin has not been identified and analyzed so far.
We do not further explore the origin of various harmonics of rotation period in the periodograms except using
the differences that clearly associate to different magnetic features.

A comparison of GLS periodograms of spectroscopically derived Sun-as-a-star quantities, 
$\Delta$RV, $log(R'_{HK})$, and
SORCE TSI and those of disk-averages of resolved observations in UV intensities, $<I_{1600}>$ and $<I_{1700}>$, 
from SDO/AIA is shown in Figure \ref{fig:13}. Interestingly, it is noted that the magnetic activity in the chromosphere 
as captured by $log(R'_{HK})$ (panel $d$) and $<I_{1600}>$ (panel $c$) do not exhibit any significant periods shorter 
than 10 days, while the other quantities of photospheric origin $\Delta$RV (panel $e$), TSI (panel $a$) or that with
significant photospheric contribution, $<I_{1700}>$, do. Now among the different magnetic
features, we see only the IN fields exhibit significant power at periods shorter than 10 days (Figures \ref{fig:11}
and \ref{fig:12}). This we speculate as an indication that the
weak IN fields are the cause of such short periodicities in $\Delta$RV and TSI, as it is known that the
footpoints of small-scale loops comprising IN fields do cause brightening in certain visible spectral bands 
such as G-band while their looping magnetic field not really reaching the chromosphere and thus not causing any
significant emissions there (see \citet{2019LRSP...16....1B} and references therein).

\section{Discussion and Conclusion} \label{sec:conclusion}
The proximity of the Sun provides us opportunities to study in detail the complex interactions
between convection and magnetic fields. Solar observations show that the structuring of 
magnetic field in terms of strengths and fluxes (sizes), which in turn determine their characteristic 
thermal and mechanical appearances, is the result of complex magnetoconvective processes \citep{2006RPPh...69..563S}.
Our understanding that such processes derive from basic physical effects in magnetohydrodynamics (MHD), 
viz. magnetic forces modifying the fluid motion, indicate that such magnetic structuring is 
expected in the photospheres of all stars like the Sun. Hence, it is hardly surprising that the
magnetic activity cause jitter in stellar RVs, and that it is mainly through the suppression of 
convective blue shifts by photopsheric magnetic fields \citep{2018ApJ...866...55C,2019Geosc...9..114C}.
However, we point out that we still lack a full understanding of the complex interplay between
convection and magnetic fields even on the Sun, and hence lack a modeling capability to account
for all the RV jitter that magnetic fields can cause \citep{2021arXiv210714291C,2023AJ....165..151N}.

In this paper, using disk-resolved images of magnetic fields and continuum intensities from SDO/HMI, 
we have attempted incorporating physical connections between the evolution of different magnetic
features on the solar surface to differentiate them better while studying their correlations with the RVs. 
Using hourly data from HMI (24 images per day), our analysis method resulted in identifying and separating 
the quiet-Sun weak internetwork magnetic fields, IN. 
Our results presented in Section \ref{subsec:f_and_fB} show that, despite its lower resolution of 1$''$,
HMI observations do capture a significant amount of IN flux. For IN fields,
high-resolution observations give a mean longitudinal flux of about 9 $\times 10^{16}$ Mx \citep{2019LRSP...16....1B},
which although is higher than HMI's detection limit of about 4.26 $\times 10^{16}$ Mx (corresponding to
a $\sigma$ of 8 G for HMI LOS measurements) our adopted 3$\sigma$ cutoff means that we
are only including IN fields with flux larger than about $10^{17}$ Mx. Hence, with a
caveat that we miss a large amount of IN flux in our study here, we are still able to identify and measure the IN
fields in HMI observations and distinguish them from the NE fields based on the above criteria
drawn from the well established physics behind the structuring of small-scale fields.
As regards the use of a flux limit to separate the IN and NE fields, although high-resolution
observations broadly agree on such a value ($\sim$ 3$\times 10^{17}$ Mx) \citep{2014ApJ...797...49G,
2020A&A...644A..86P}, we note that it is not a strict limit as these fields have
significantly broad distributions with observed peak fluxes dependent on the resolution of observations
\citep{2013SoPh..283..273Z} and hence it is always possible that there exist kG NE elements with a slightly
smaller flux and vice versa for IN elements, because intermediate states of splitting and merging are a
common occurrence in the dynamics of NE \citep{1997ApJ...487..424S} and IN \citep{2021A&A...647A.182C} fields .

We note that the differing signatures of different magnetic features carry important information on the dynamical
connections between the evolution of different magnetic flux concentrations and their interactions with
convective and other large-scale flows on the Sun. In the context of RV variations,
we particularly note the characteristic short time-scale fluctuations of IN fields, which while at the
same time changing relatively very little over the longer cyclic time-scale, compared to other features. This latter
aspect of IN fields is also well appreciated in the solar literature \citep{2019LRSP...16....1B}.
We have shown that these two features of IN fields potentially introduce such time-scales in the RV variations
too (cf. Figures \ref{fig:6} and \ref{fig:12}), while also confirming the previously reported dominant correlations
between other magnetic structures (plages and spots) and RVs. 
And, as shown in Sections \ref{subsec:corrs} and \ref{subsec:periodogram}, we are also able to
distinguish the differing signatures of IN and NE fields in the HARPS-N RV variations (Figures \ref{fig:11} and \ref{fig:12}).
Such a contribution from IN fields is expected because flux contained in them is significant (cf.
Figure \ref{fig:3}) and moreover they dominate the total flux during the solar minimum period as evident in the relative
variations of $(fB)_{m}$ shown in Figure \ref{fig:4}. The IN fields changing relatively little over the solar cycle
perhaps cause a constant background signal in $\Delta$RV. 
We stress that the relative constancy of IN on longer time scales, thus, is important to correctly determine the 
long-term baseline in the RV fluctuations due to magnetism: in the absence of full coverage of a solar minimum period, a
mean-subtracted RV would be biased by the strongly varying contributions from
active region fields and hence would cause an offset as seen in Figure 7.
We further point out that weak background fluctuations from the IN fields
on the long cycle time-scale is of significant consequence for the following reasons:
(i) we still do not understand the origin of the weak fields (holding a large fraction of the total flux) on the Sun
\citep{2008ApJ...672.1237L,2019LRSP...16....1B}, although simulation studies show the possible 
opertaion of the so called local (small-scale) dynamos with wider implications for stellar
magnetism \citep{2023SSRv..219...36R,2023NatAs...7..662W}, and
(ii) such fields may be of greater relevance in other stars that possibly maintain
them much more efficiently and hence extreme precision measurements of RVs of these
stars should provide pathways to explore the existence of these fields and the underlying
dynamo mechanisms.
We note that at 1$''$ resolution SDO/HMI captures only about 1/3rd of flux in IN \citep{2006ASPC..358...42K,2013SoPh..283..273Z}, 
and hence our inferences on the contributions of IN to RV variations are likely to be much lower than the actual ones. 
We have also derived indications that correlations
between fill-factors and average unsigned magnetic fluxes of different magnetic features themselves 
may play a role in the differences between the correlations of area fill-factors and average unsigned magnetic fluxes
with RVs.

Further, we point out that the correlated variation between spots and IN fields, presented and discussed
in Section \ref{subsec:f_and_fB} (see the insets in Figures \ref{fig:2} and \ref{fig:3}),
is related to the nature of decay of active region flux, wherein the larger content of spots
decay directly to the weak IN on significantly shorter time-scales than that of re-organisation of such flux
into more uniform supergranular NE fields. Hence, if NE indeed receives a major supply from decaying active
regions then there will be a significant delay between the rise of spot flux and that in NE, depending on
the time-scales involved in the dispersal of all the decaying flux routed via the IN. Noting that, since the flux
accumulating in it gets intensified and brighter, the NE is a significant contributor
to the total solar irradiance (TSI) enhancements that compensates for reduction due to dark spots, the relationships
between temporal variations of feature-specific $f_{m}$ and $(fB)_{m}$ that we see in Figures \ref{fig:2} and \ref{fig:3}
are likely behind those seen in the brightness variations contributed by faculae, network and spots \citep{yeoetal2020}.

With the aim of exploring and deriving additional diagnostics from UV emissions observed by
the SDO/AIA, we have experimented with disk-averaged emission intensities at wavelengths 1600 \r{A} and 1700 \r{A}, 
$<I_{1600}>$ and $<I_{1700}>$. These UV emissions are well known to capture the underlying magnetic flux in the 
photospheric and chromospheric heights, although the exact mechanisms of heating that cause emissions may differ depending 
on the height ranges and the spectral content of these two wavelength bands \citep{2001A&A...379.1052K}.
We have studied correlations of these UV emissions with different magnetic features, especially comparing them with those
of spectroscopically derived Ca II K index $log(R'_{HK})$ from HARPS-N. While we find that the disk-averaged UV intensities
perform, in general, equally well as a chromospheric proxy for the magnetic flux behind the RV variations as $log(R'_{HK})$, 
we also show that the UV intensities remain linearly correlated with plage magnetic flux at high activity levels, while
the Ca II H-K emission indices tend to saturate. Within the time period of data used in this work, 
we observe only a slight saturation in $log(R'_{HK})$ against plage fields ($c.f.$, Figures \ref{fig:9} and \ref{fig:10}).
However, at cycle maximum activity levels when the disk areas of plage fields peak, the saturation of chromospheric Ca-II K 
emissions is well known and studied on the Sun (see, for example \citet{2009A&A...497..273L}). Such a phenomenon is also
well observed in a large number of other Sun-like stars (see for example \citet{2022A&A...662A..41R}). Hence, our finding 
of stronger correlations between UV intensities ($<I_{1600}>$ and $<I_{1700}>$) and plage fields (Figures \ref{fig:9} and 
\ref{fig:10}) point to a better utility of these emission measures in capturing contributions of plage fields to RVs, 
especially at high activity levels or in highly active stars, which saturate in chromospheric Ca-II K emissions.

Lastly, through the generalised Lomb-Scargle (GLS) periodogram analysis \citep{2009A&A...496..577Z}, 
we have identified short term periodicities possibly arising from the dynamics of our newly characterised internetwork (IN) 
fields and have related them to such periodicities seen in HARPS-N radial velocity fluctuations ($\Delta$RV).
We have also noted that the power peaks at 9 - 10 day periods coincide with the third harmonic of the primary 
rotation period (27 - 30 days). However, since only the weak internetwork fields (IN) exhibit  
significant peaks at periods shorter than 10 days we have speculated that the higher (third) harmonics is possibly 
related to their spatial (latitudinal) distribution and time-evolution, which interfere with that due to
differential rotation. Another interesting feature that favors such an origin, related to the physical distribution
of IN fields, of periods close to the third harmonic is 
the difference between periodograms of photospheric and chromospheric quantities: significant short-term periodicities ($<$ 10 days)
appear only in the periodograms of photospheric observables, HARPS-N RV, SORCE TSI and slightly in $<I_{1700}>$, 
but not in the chromospheric ones, $log(R'_{HK})$ and $<I_{1600}>$.
Since only the IN fields, among the different magnetic features, exhibit periodicities shorter than 10 days 
(Figures \ref{fig:11} and \ref{fig:12}), we conclude that the weak IN fields are the cause of such variations
in $\Delta$RV. This conclusion is strengthened further as the TSI is known to receive contributions from the
foot points of small-scale loops comprising IN fields, which do cause brightening in certain visible spectral bands
such as G-band while their looping magnetic field not really reaching the chromosphere and thus not causing any
significant emissions there (see \citet{2019LRSP...16....1B} and references therein). 
Further, given that the origin of all the flux in IN and NE, which is more than 50\% of the total flux on the Sun 
during activity maximum, is not yet fully understood \citep{2008ApJ...672.1237L,2019LRSP...16....1B}, dynamics of 
such magnetic fields in other stars, depending on their relative flux content with respect to other well known 
features such as plages and spots, may play important roles in the variations of stellar RVs. A careful survey
of precision RVs of Sun-like stars with varying activity levels will thus be importance not only for exoplanet
studies but also to understand stellar magnetism.

\section{Acknowledgements} 
The HMI and AIA data used are courtesy of NASA/SDO and the HMI and AIA science teams.
Data preparation and processing have utilised the Data Record Management System (DRMS) software at the 
Joint Science Operations Center (JSOC) for NASA/SDO at Stanford University.
HARPS-N solar data used are courtesy of the Geneva Observatory, the Center for Astrophysics in
Cambridge (Massachusetts), the Universities of St. Andrews and Edinburgh, the Queen's University
Belfast, the UK Astronomy Technology Centre, and the Italian Istituto Nazionale di Astrofisica.
Data intensive computations in this work have utilised the High-Performance Computing 
facility at the Indian Institute of Astrophysics. A.S. is supported by INSPIRE Fellowship from the 
Department of Science and Technology (DST), Government of India. S.P.R. acknowledges support
from the Science and Engineering Research Board (SERB, Government of India) grant CRG/2019/003786.
We acknowledge communications with Dr. Xavier Dumusque on the latest version, ESPRESSO DRS 2.3.5, 
of HARPS-N data pipeline for the RV and Ca II K data used in this work. 
We acknowledge the use of free packages of Python software. We thank our colleagues Harsh Mathur, T. Sivarani,
Athira Unni, Swastik Chowbay and Ravinder Banyal at the Indian Institute of Astrophysics for useful discussions. 

\noindent Author Contributions:\\
S.P.R. contributed to the conception and design of the work reported in this paper.
A.S. collected the data, developed the numerical codes and carried out the computations.
Both the authors jointly worked on analysing the calculations and intepreting the results.
A first draft of the paper was written by A.S. S.P.R. worked on correcting and improving
the presentation and prepared the final draft.


\bibliography{msbib}{}
\bibliographystyle{aasjournal}

\begin{figure*}[!htbp]
        \gridline{\fig{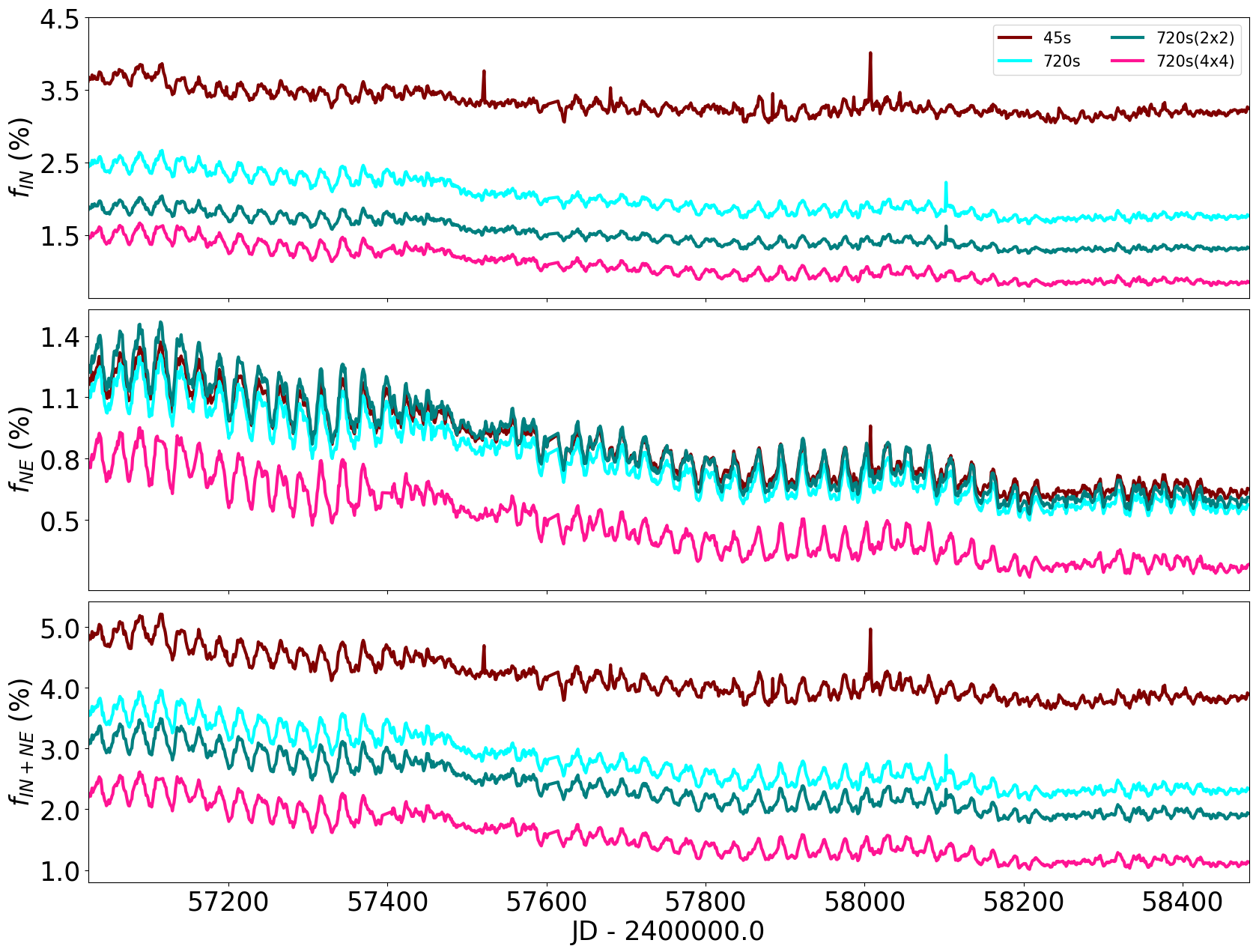}{0.8\textwidth}{}
          }
        \caption{Comparison of fill-factors $f_{IN}$, $f_{NE}$ and total network ($f_{ntwk} = f_{IN}+f_{NE}$) for 45sec
cadence data (without binning), and 720 sec cadence datasets (2x2, 4x4 binning and without binning). See Section \ref{subsec:f_and_fB}
        for a discussion and summary of these results.}
        \label{fig:14}
\end{figure*}

\appendix
Here, we present results from tests carried out comparing 45 sec and 720 sec exposures HMI data, including different spatial
binnings ($2\times2$ and $4\times4$), to identify the cause of differences in fill-factors of network magnetic fields,
$f_{IN}+f_{NE}$, which is the sum of those of weak IN and strong NE fields (sum of top two panels of Figure \ref{fig:2}), between our
work and that of \citet{milbourne2021estimating,haywood2022unsigned} ($f_{ntwk}$ in Figure 2 of \citet{milbourne2021estimating}
or Figure 4 of \citet{haywood2022unsigned}). Figure \ref{fig:14} shows the results obtained from our 
repeat of analysis described in Section \ref{sec:sub_method} using the 720-sec exposure data from HMI with different levels
of spatial binning. \\
\begin{table*}[!h]
  \centering
  \small
  \makebox[\linewidth]{%
    \begin{tabular}{| *{9}{c|} }
    \hline
      Author & Cadence &Spatial Binning & Spot(\%) & Faculae(\%) & Plage(\%) & IN+NE(\%) & NE(\%) & IN(\%) \\

      & & & & (Plage+IN+NE) & & & &  \\
      \hline\hline
      \multicolumn{9}{|c|}{Date: 10 November 2011, 00:01:30 UTC}\\
       \hline\hline
       \citet{haywood2016sun}(fig-3) & 45s & No & 0.4 & 9 & NA & NA & NA & NA\\
       \hline
       This Work & 45s & No & 0.41 & 9.03 & 4.31 & 4.72 & 1.15 & 3.57 \\
       \hline
       \hline
       \multicolumn{9}{|c|}{Date: 28 November 2015, 20:00:00 UTC}\\
       \hline\hline
       \citet{milbourne2019harps}& & & & & & & & \\
       \citet{haywood2022unsigned}(fig-1) & 720s & Not Known & 0.03 & 3.25 & 1.59 & 1.66 & NA & NA  \\
        \citet{milbourne2021estimating}& & & & & & & & \\
       \hline
       This Work & 720s & No & 0.03 & 4.82 & 1.80 & 3.02 & 0.87 & 2.15  \\
       & & 2x2 & 0.03 & 4.30 & 1.71 & 2.59 & 0.96 & 1.63 \\
       & & 4x4 & 0.03 & 3.38 & 1.63 & 1.75 & 0.56 & 1.19 \\
       \hline
       This Work & 45s & No & 0.03 & 5.98 & 1.81 & 4.17 & 0.94 & 3.23 \\
       & & 2x2 & 0.03 & 4.92 & 1.70 & 3.22 & 0.99 & 2.23  \\
       & & 4x4 & 0.03 & 3.43 & 1.58 & 1.85 & 0.57 & 1.28 \\
       \hline

    \end{tabular}%
  }
  \caption{Fill-Factor Comparison with earlier authors for two dates}
  \label{Table1}
\end{table*}

These results for fill-factors have to be compared with those presented in Figure \ref{fig:2} for 45 sec data.
Using the four sets of data, we find that the fill-factors of large magnetized regions
like spots and plage are not affected, as expected, whereas those of the small structures (IN, NE, and IN+NE)
differ significantly for the different cadences and spatial binning used. We find that most of the previous studies have missed 
capturing a significant amount of this small-scale magnetic fields because of their use of 720sec cadence dataset with a likely 
further 4x4 binning. A discussion and summary of these results are given in Section \ref{subsec:f_and_fB}.


To compare specific numbers for fill-factors with the previous studies, especially with those of
\citet{haywood2016sun} and \citet{haywood2022unsigned}, we tabulated results for the same dates (given in their papers)
in Table \ref{Table1}. \citet{haywood2016sun} use the
original 45sec cadence LOS magnetogram data and explicitly mention the fill factors of sunspots and faculae (all non-spot fields),
observed on 2011 November 10 at 00:01:30 UTC (in Figure 3 of \citet{haywood2016sun}). We find that our result matches exactly
with that of \citet{haywood2016sun}. In a similar way, we have tested our result for the dataset taken
on 2015 November 28 at 20:00:00 UTC, to compare with Figure 1 of \citet{haywood2022unsigned}.
We find that our test results match the best with previous ones only after performing a 4x4 binning on the 720sec HMI LOS magnetogram.

\end{document}